# Observation of optical chaotic solitons and modulated subharmonic route to chaos in mode-locked laser

Huiyu Kang[1,†], Anran Zhou[1,†], Ying Zhang[1], Xiuqi Wu[1], Bo Yuan[1], Junsong Peng[1,2,3,*], Christophe Finot[4], Sonia Boscolo[5], Heping Zeng[1,2,3,6**]

[1]State Key Laboratory of Precision Spectroscopy, East China Normal University, Shanghai 200062, China

[2]Collaborative Innovation Center of Extreme Optics, Shanxi University, Taiyuan, Shanxi 030006, China

[3]Chongqing Key Laboratory of Precision Optics, Chongqing Institute of East China Normal University, Chongqing 401120, China

[4]Laboratoire Interdisciplinaire Carnot de Bourgogne, UMR 6303 CNRS—Université de Bourgogne Franche-Comté, F-21078 Dijon Cedex, France

[5]Aston Institute of Photonic Technologies, Aston University, Birmingham B4 7ET, United Kingdom

[6]Chongqing Institute for Brain and Intelligence, Guangyang Bay Laboratory, Chongqing, 400064 China

† These authors contributed equally to this work

*jspeng@lps.ecnu.edu.cn  ** hpzeng@phy.ecnu.edu.cn

We reveal a new scenario for the transition of solitons to chaos in a mode-locked fiber laser: the modulated subharmonic route. Its universality is confirmed in two different laser configurations, namely, a figure-of-eight and a ring laser. Numerical simulations of the laser models agree well with the experiments. The modulated subharmonic route to chaos could stimulate parallel research in many nonlinear physical systems.

Solitons, i.e., stable self-localized wave packets, and chaos, i.e., unpredictable behavior appearing irregularly due to a high sensitivity to small changes in initial conditions, are two complementary manifestations of nonlinearity. Interestingly, several theoretical studies have shown that solitons can behave chaotically in perturbed systems [1-5]. While the experimental observation of chaotic solitons has so far been restricted to spin wave systems [6-9], the chaos engendered by optical solitons is also of great interest. Indeed, in 1975, Haken discovered an analogy between the Maxwell–Bloch equations for lasers and the Lorenz equations for chaos [10], opening the study of laser chaos. Lasers can also generate solitons, as described by the generalized standard nonlinear Schrödinger equation (GNLSE) [11] [12]. Therefore, the chaotic behavior of solitons can extend laser chaos to the framework of the GNLSE. The significance of such an extension is twofold. Firstly, the generation of laser chaos within the Maxwell-Block equations requires externally injected signals [13], otherwise, the required pump power is too high [14]. By contrast, soliton chaos can be generated in a free-running mode-locked laser, as theoretically predicted in [5]. It is worth to note here that the nonlinear coupling between two polarized modes can also trigger chaos in a laser without requiring external signals [15]. Secondly, vastly different physical systems follow three well-defined universal routes to chaos, namely, the Ruelle-Takens scenario (quasiperiodicity route) [16], the Feigenbaum scenario (period-doubling route) [17] and the Pomeau-Manneville scenario (intermittency route) [18]. The dynamics of these systems are normally governed by the Lorenz equations or one-dimensional maps. As the GNLSE is rather different from these models, new mechanisms of transition from regular to chaotic dynamics can be anticipated in GNLSE-governed systems. Observing new, relatively universal routes to chaos in real systems represents an issue of fundamental importance in nonlinear science [19-26].

Besides its significance in fundamental science, soliton chaos also holds great potential for applications. It is inherently localized and, therefore, could extend the application areas of chaos theory. Additionally, chaotic solitons have a broad optical spectrum, which can make more channels available for parallel ranging than chaotic microcombs generated through modulation instability [27, 28].

Yet, despite the importance of soliton chaos and its theoretical prediction in a mode-locked laser in 2005 [5], experimental studies of optical chaotic solitons remain rare because the ultrafast soliton dynamics are far beyond the resolution of state-of-the-art electronic devices. Previous experimental works on chaotic solitons in mode-locked lasers suffer from this difficulty and, thus, the chaotic behaviour of solitons, which require simultaneous high-resolution and real-time detection of the solitons, could not be confirmed [29]. Similar difficulty also exists in characterizing chaos in Kerr resonators [30]. A very recent study [31] has reported a systematic experimental investigation of the transition of solitons to chaos in a mode-locked laser following the scenario of cascaded short- and long-period pulsations through radio-frequency (rf) spectrum measurements and Lyapunov exponent analysis.

In this Letter, by using real-time measurements [32-35], we demonstrate chaotic dynamics of solitons in a mode-locked fiber laser. The chaotic soliton behavior is unambiguously confirmed. A new route to chaos – dubbed 'modulated subharmonic route' - is revealed, which is distinctly different from the subharmonic route reported by works in other fields [21, 22]. Indeed, in this scenario, the transition from solitons to chaos occurs through a laser state which is characterized by self-modulation of sub-harmonically synchronized breather oscillations [36]. The universality of these observations is demonstrated in two laser configurations.

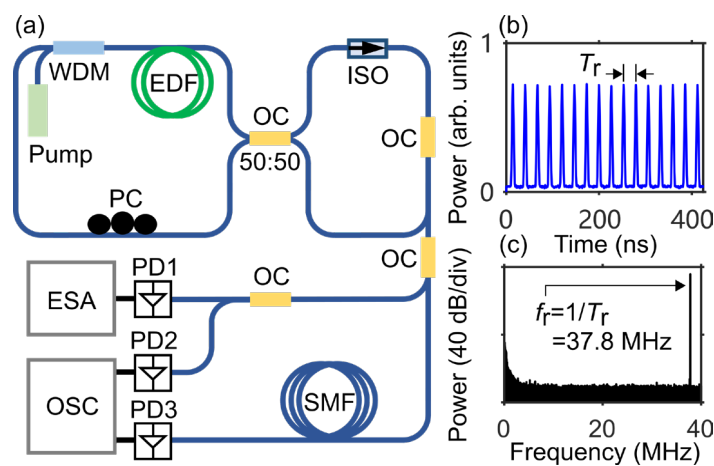

Fig. 1 (a) Figure-of-eight fiber laser setup. WDM: wavelength division multiplexer, EDF: erbium-doped fiber, OC: optical coupler, ISO: isolator, PC: polarization controller, SMF: single-mode fiber, PD: photodetector, ESA: electrical spectrum analyzer, OSC: oscilloscope. The left loop includes 1.22-m EDF and 3.14-m SMF, while the right loop consists of 1-m SMF. (b) Soliton train observed on the OSC. (c) Corresponding rf spectrum.

The figure-of-eight laser consists of a passive unidirectional ring and an active bidirectional ring connected via a 3-dB fiber coupler (Fig. 1). Unidirectionality of the light in the passive ring is realized via an optical isolator. The active ring contains a short section of erbium-doped fiber asymmetrically placed in the loop, pumped by a laser diode through a wavelength-division multiplexer. This creates a power imbalance between the counterpropagating light beams in the loop, which induces a differential phase, and consequently a power-dependent reflectivity that mimics the action of a saturable absorber, promoting pulse generation in the laser [37]. A polarization controller is employed to finely adjust the phase difference between the two propagation directions. To demonstrate that the chaotic dynamics observed do not depend on the particular laser configuration being considered, we have also used a ring-cavity laser using the nonlinear polarization evolution (NPE) mode-locking mechanism [38], similar to the one studied in Ref. [36] (Supplemental Material Fig. S1).

The output signal from the laser is detected by an electrical spectrum analyzer (ESA) to study its rf power spectrum. To capture the fast evolution of the circulating pulses over consecutive cavity roundtrips, we employ the real-time measurement technique known as time stretch [32, 35, 39, 40]. It utilizes dispersion to stretch a pulse so that the different frequency components of the pulse can be separated in time, thereby enabling optical spectrum measurements with an oscilloscope, which works significantly faster than an optical spectrum analyzer.

By increasing the pump strength above the threshold, mode locking in the soliton generation regime can be observed [37], as evidenced by the periodic pulse train with constant energy measured by the oscilloscope (Fig. 1b). The repetition frequency of the pulse train, measured by the ESA, is $f_r = 1/T_r = c/nL = 37.8$ MHz (Fig. 1c), where $T_r$ is the pulse spacing, $L$ is the cavity length, $n$ is the refractive index of the fiber and $c$ is the speed of light. Different operating states can be successively accessed by tuning the pump current only, as revealed by the rf spectrum measurements shown in Fig. 2a. Representative cross-sections of Fig. 2a are plotted in Fig. 2b, and the corresponding energy evolutions are shown in the Supplemental Material (Fig. S2). The soliton regime, existing in our laser at the pump currents around 135 mA, does not feature any signals in Fig. 2a as we have used a reduced frequency range (3 to 9 MHz) for better view. The same spectrum measurements are shown in Fig. S3 of the Supplemental Material across a wider frequency span from 0 to $f_r$.

Increase of the pump current above 135.2 mA leads to the excitation of breathing solitons, as evidenced by the appearance of a signal (the breathing frequency $f_b$) around 6 MHz. The transition from stationary to breathing soliton correlates to the ubiquitous phenomenon known as Hopf bifurcation, which has been observed in various fields including opto-mechanics [41-43] and fluids [44, 45] and recently drawn considerable attention also in laser and Kerr resonators [36, 46-56]. The breathing frequency can be either locked or unlocked to the cavity repetition rate, resulting in a subharmonic or non-subharmonic breather state, respectively, as indicated in Fig. 2a, where the subharmonic state features a rational frequency ratio (or winding number) of $f_b/f_r = 1/6$. The distinctiveness of subharmonic versus non-subharmonic breather states has been studied in [51], showing that the subharmonic states are much more stable than the non-subharmonic ones, which is reflected in a high-quality breathing frequency with a narrow linewidth and a high signal-to-noise ratio. Further pump current increase (to around 152 mA) causes a transition from the subharmonic regime to the so called 'modulated subharmonic' regime [36], characterized by the appearance of two symmetric sidebands around the subharmonic breathing frequency, which manifest themselves as additional temporal modulations on the pulse energy (Fig. S2 of Supplemental Material). But these sidebands drift under pump current increase; they have a noisy structure and are not stable (Supplemental Material, Figs. S4 and S5). Such instabilities eventually lead to the onset of chaos, evidenced by a significantly broadened rf spectrum (Fig. 2a and Fig. 2b). To validate this chaotic behavior, we have further investigated Poincaré maps, Lyapunov exponents and correlation dimensions.

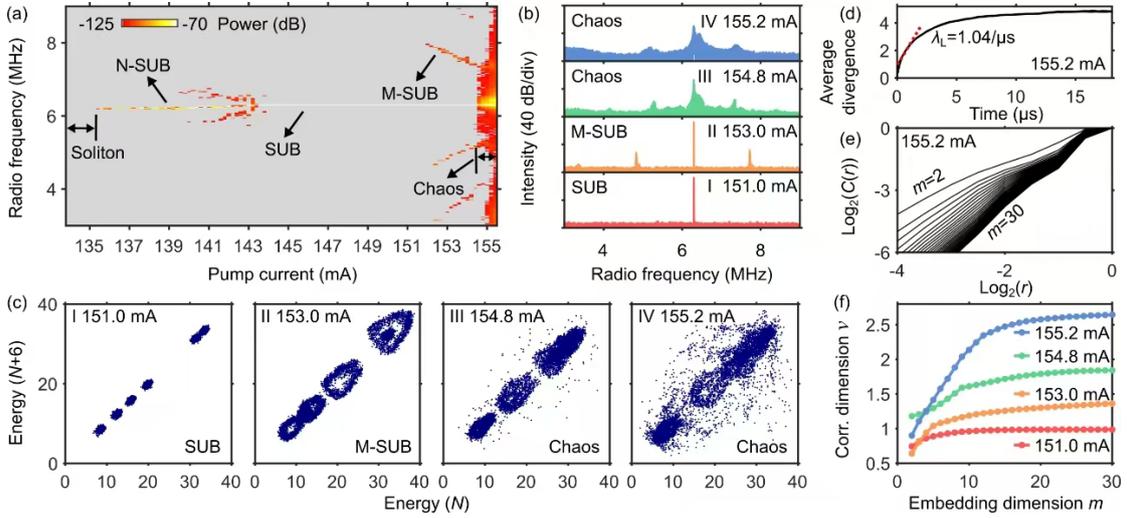

**Fig. 2** Experimental observation of modulated subharmonic route to chaos. (a) Rf spectrum of the laser output as a function of the pump current, showing successive phase transitions from stationary solitons to non-subharmonic (N-SUB), subharmonic (SUB), modulated subharmonic (M-SUB) breathing solitons and to chaos. (b) Representative rf spectra of subharmonic (151 mA), modulated subharmonic (153 mA), and chaotic (154.8 and 155.2 mA) states. (c) Corresponding Poincaré sections, showing the phase portraits of the pulse energy at roundtrip $N$ versus the energy at roundtrip ($N$+6). (d) Maximum Lyapunov exponent analysis of the chaotic state in panel IV of (c). An average of the divergence over all nearby initial trajectories is fitted by $e^{\lambda_L t}$, yielding the exponent fitting result $\lambda_L =1.04/\mu s$. (e) Grassberger-Procaccia algorithm for the chaotic state in panel IV of (c), showing plots of the correlation integral ($C(r)$) versus the sphere radius ($r$) for varying embedding dimension ($m$, from 2 to 30). The exponent of the power-law dependence of $C(r)$ at small $r$ provides an estimate of the correlation dimension. (f) Correlation dimension versus embedding dimension for the four dynamic states in (b).

Poincaré maps or first-return maps are often used to analyze the chaotic dynamics of dissipative systems by projecting the system's trajectories onto a lower dimensional space [57-59]. Figure 2c shows the sequence of Poincaré sections corresponding to the laser operating states depicted in Fig. 2b. The subharmonic state exhibits six (fixed) points (panel I), which transform to open loops in the modulated subharmonic state (panel II). These isolated loops are subsequently connected by many points (panel III), resulting in an expanded phase space, which marks the onset of chaos. The fully chaotic regime is then entered (panel IV). Importantly, the Poincaré section in panel III displays a periodic solution indicated by the open loops, coexisting with a certain amount of chaotic motion. This is also evident from the corresponding rf signal (Fig. 2b (III)), where the single peak indicates periodicity while the broad background is a manifestation of chaos. These intriguing dynamics are a distinctive signature of the phenomenon of chaotic resonance predicted in [5].

The Lyapunov characteristic exponent ($\lambda_L$) of a dynamical system, quantifying the divergence rate of any two initially nearby trajectories in the phase space, provides an estimate of the amount of chaos in the system [59, 60], where a positive $\lambda_L$ signals chaos. The fully chaotic state analyzed in Fig. 2d, shows an exponential divergence with $\lambda_L$=1.04/µs. The calculation of the Lyapunov exponent for the other dynamic states is shown in the Supplemental Material (Fig. S6), where $\lambda_L$=0.399/µs for the chaotic resonance state and a negative $\lambda_L$ characterizes the subharmonic state. Although $\lambda_L$ should be zero theoretically for the modulated subharmonic state, it is in fact positive because of noise in the measurements, but its value is much smaller (0.045/µs) than those of the chaotic states.

The correlation dimension ($\nu$) is another metric for characterizing chaos and can be obtained from the time series produced by a dynamical system using the method of Grassberger and Procaccia [61]. It can be used to distinguish between stochastic noise and deterministic chaos, as in the case of noise-perturbed systems it increases with increasing embedding dimension $m$, whilst for deterministic chaos it saturates with $m$. Figure 2e shows plots of the correlation integral $C(r)$ against the sphere radius $r$ on a doubly logarithmic scale for different values of $m$ for the chaotic system (panel IV of Fig. 2c; the corresponding plots for other states are given in Fig. S7 of the Supplemental Material), clearly highlighting the saturation of $\nu$ (estimated from the small $r$-behavior of $C(r)$). Such a saturation is further evidenced by Fig. 2f, which shows that all the four states have a saturated $\nu$. As the correlation dimension reflects the complexity of the associated attractor, it follows the trends shown in Fig. 2c, with the fully chaotic state featuring the largest value of 2.6. For comparison, we have also calculated the correlation dimension for noise-like mode locking, which is unsaturated (Supplemental Material, Fig. S8).

It is noteworthy that the pulse energy used to construct the phase portraits in Fig. 2c is obtained from the optical spectrum of each individual pulse measured by time stretch (Fig. 1, PD3). By contrast, constructing the phase portraits from direct temporal intensity measurements (Fig. 1, PD2), conceals the detailed structure of chaos due to the limited resolution of the photodetector, which cannot resolve the soliton duration around 665 fs (Supplemental Material, Figs. S9 and S10). This corroborates the key role played by fast detection techniques in exploring optical soliton chaos. The modulated subharmonic route to chaos that we have observed in the ring-cavity laser is illustrated in the Supplemental Material (Figs. S11 and S12 show the experimental and numerical results, respectively).

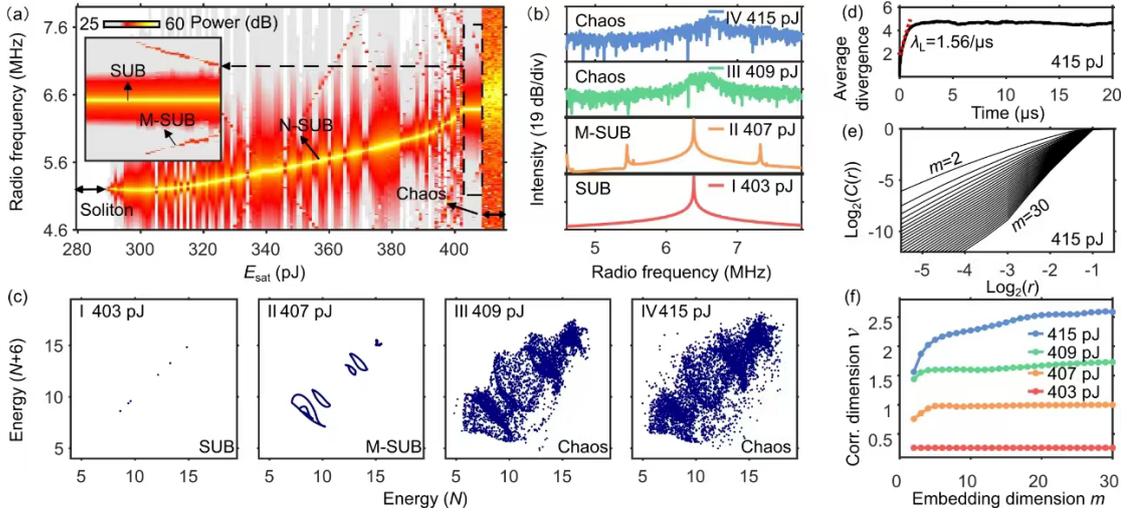

**Fig. 3** Numerical simulation of modulated subharmonic route to chaos. (a) Rf spectral intensity as a function of the gain saturation energy. The region of modulated subharmonic dynamics is magnified in the inset. (b) Representative rf spectra of subharmonic (403 pJ), modulated subharmonic (407 pJ) and chaotic (409 and 415 pJ) states. (c) Corresponding Poincaré sections. (d) Maximum Lyapunov exponent analysis of the chaotic state in panel IV of (c), yielding the exponent fitting result $\lambda_L = 1.56/\mu s$. (e) Grassberger-Procaccia algorithm for the chaotic state in panel IV of (c), showing plots of the correlation integral ($C(r)$) versus $r$ for varying embedding dimension. (f) Correlation dimension versus embedding dimension for the four dynamic states in (b).

We have performed numerical simulations based on a lumped model which models each part of the laser cavity separately. Pulse propagation in the optical fibers is modeled by the GNLSE:[62, 63]

$$\psi_z = -\frac{i\beta_2}{2}\psi_{tt} + i\gamma|\psi|^2\psi + \frac{g}{2}(\psi + \frac{1}{\Omega^2}\psi_{tt})$$

where $\psi = \psi(z,t)$ is the slowly varying complex pulse envelope in the comoving frame, moving at the group velocity $v_g = d\beta/d\omega$ along the coordinate $z$, where $t = t' - z/v_g$ is the retarded time (with $t'$ the present time). $\beta_2$ and $\gamma$ are the group-velocity dispersion and Kerr nonlinearity coefficients, respectively. The dissipative terms represent linear gain and a parabolic approximation to the gain profile with the bandwidth $\Omega$. The gain is saturated according to $g(z) = g_0/[1 + E(z)/E_{sat}]$, where $g_0$ is the small-signal gain (non-zero only for the gain fiber), $E(z) = \int |\psi|^2 dt$ is the pulse energy, and $E_{sat}$ is the gain saturation energy. The latter can be used to change the pulse energy, thus playing a similar role to the pump current in the experiment. The parameters used in the numerical model are provided in the Supplemental Material (Table 1). In the case of the NPE mode-locked laser, we have modelled NPE technique by an instantaneous and monotonous nonlinear transfer function for the field amplitude [36].

The evolution of the rf spectral intensity with the gain saturation energy shown in Fig. 3a clearly highlights five successive phases of the laser dynamics, namely, stationary solitons, non-subharmonic, subharmonic and modulated subharmonic (also magnified in the inset) breathing solitons, and chaos, in a good agreement with the experiments (Fig. 2a). Representative rf spectra and the associated phase diagrams are plotted in Figs. 3b and 3c, respectively. A video of the evolving phase diagrams is provided in the Supplemental Material. The chaotic behavior of the system is confirmed by a positive maximum Lyapunov coefficient (Fig. 3d) and a saturated correlation dimension (Fig. 3e). Figure 3f evidences that the correlation dimension of the chaotic attractor increases with the gain saturation energy (green and blue curves) and is much larger than that of the fixed points (red curve) or limit cycles (orange curve). Figure 4 shows results of the characterization of the chaos in terms of correlation dimension, indicating good agreement between the experiments and the numerical simulations.

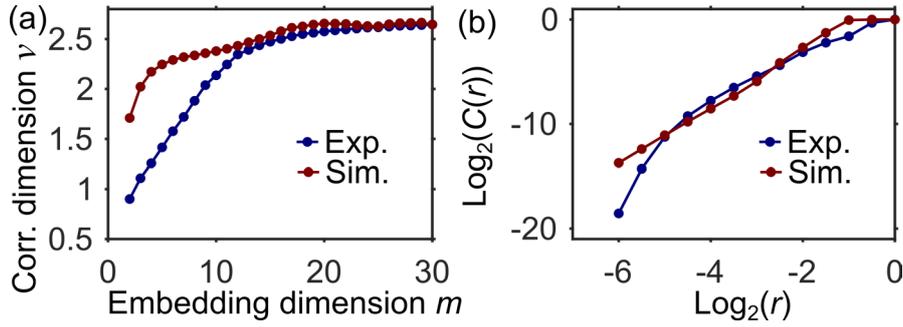

Fig. 4 (a) Correlation dimension versus embedding dimension of the chaotic state (panels IV of Figs. 2(b) and 3(b)). (b) Correlation integral versus sphere radius for $m = 30$.

Besides the new route to chaos discussed above, by setting the laser to a different polarization state in the experiment, or equivalently, by changing the loss in the model, we have also found that solitons in our laser can experience transition to chaos following the subharmonic route [21, 22] (Figs. S13 and S14 of the Supplemental Material), or through a non-subharmonic breathing soliton state ('non-subharmonic route'; Supplemental Material, Figs. S15 and S16). It is worth noting that, although Fig. 2a may suggest that chaos can be observed in the laser by simply increasing the pump current from the soliton state, this scenario depends crucially on the angular parameters of the polarization controller. Typically, in fact, increasing the pump current from the soliton state does not generate chaos, and other laser regimes are visited (Supplemental Material, Fig. S17).

**Conclusion**
The experimental research on chaotic solitons is still at its infancy mainly because measuring ultrafast pulse dynamics with high resolution remains a challenging task. Here, using real-time detection techniques, we have revealed a new route from solitons to chaos through modulated subharmonic breather oscillations in both experiments and numerical simulations with lasers using two different configurations, thereby proving its ubiquity in lasers. The modulated subharmonic route to chaos is expected to be observed in many systems since the GNLSE represent a universal model describing numerous physical systems. Besides, the number of degree of freedom increases with the number of solitons, thus a laser emitting multiple solitons could open the possibility to explore hyperchaos [64, 65], and synchronization of chaos using multiple solitons [66, 67].

**Acknowledgments**

We acknowledge support from Innovation Program for Quantum Science and Technology (2023ZD0301000), the National Natural Science Fund of China (11621404, 11561121003, 11727812, 12074122, 62475073 and 12434018), Shanghai Natural Science Foundation (23ZR1419000), Engineering and Physical Sciences Research Council (EP/X019241/1), NATO Science for Peace and Security Programme (G6137).

# Observation of optical chaotic solitons and modulated subharmonic route to chaos in mode-locked laser: Supplementary Materials

Huiyu Kang[1,†], Anran Zhou[1,†], Ying Zhang[1], Xiuqi Wu[1], Bo Yuan[1], Junsong Peng[1,2,3,*], Christophe Finot[4], Sonia Boscolo[5], Heping Zeng[1,2,3,6**]

[1]State Key Laboratory of Precision Spectroscopy, East China Normal University, Shanghai 200062, China

[2]Collaborative Innovation Center of Extreme Optics, Shanxi University, Taiyuan, Shanxi 030006, China

[3]Chongqing Key Laboratory of Precision Optics, Chongqing Institute of East China Normal University, Chongqing 401120, China

[4]Laboratoire Interdisciplinaire Carnot de Bourgogne, UMR 6303 CNRS—Université de Bourgogne Franche-Comté, F-21078 Dijon Cedex, France

[5]Aston Institute of Photonic Technologies, Aston University, Birmingham B4 7ET, United Kingdom

[6]Chongqing Institute for Brain and Intelligence, Guangyang Bay Laboratory, Chongqing, 400064 China

† These authors contributed equally to this work

*jspeng@lps.ecnu.edu.cn  ** hpzeng@phy.ecnu.edu.cn

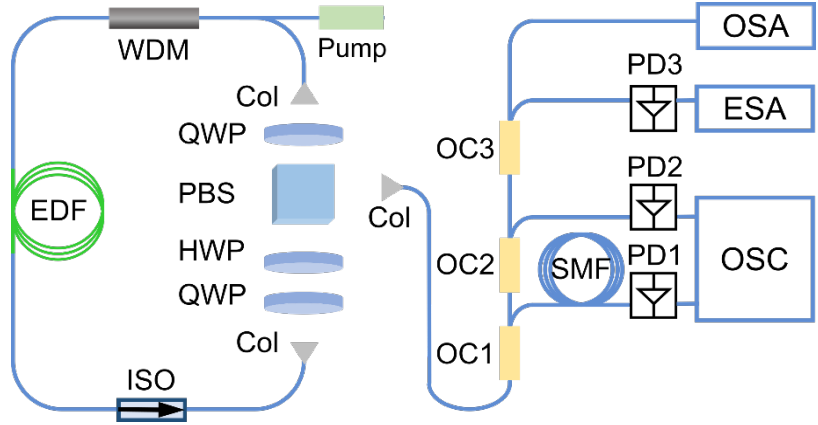

Fig. S1. Ring-cavity laser setup[1]. WDM: wavelength division multiplexer, EDF: erbium-doped fiber, ISO: isolator, QWP: quarter-wave plate, HWP: half-wave plate, PBS, polarization beam splitter, OC: optical coupler, PD: photodetector, OSA: optical spectrum analyzer, ESA: electrical spectrum analyzer, OSC: oscilloscope. The laser is mode locked by an effective saturable absorber based on the nonlinear polarization evolution (NPE) effect.

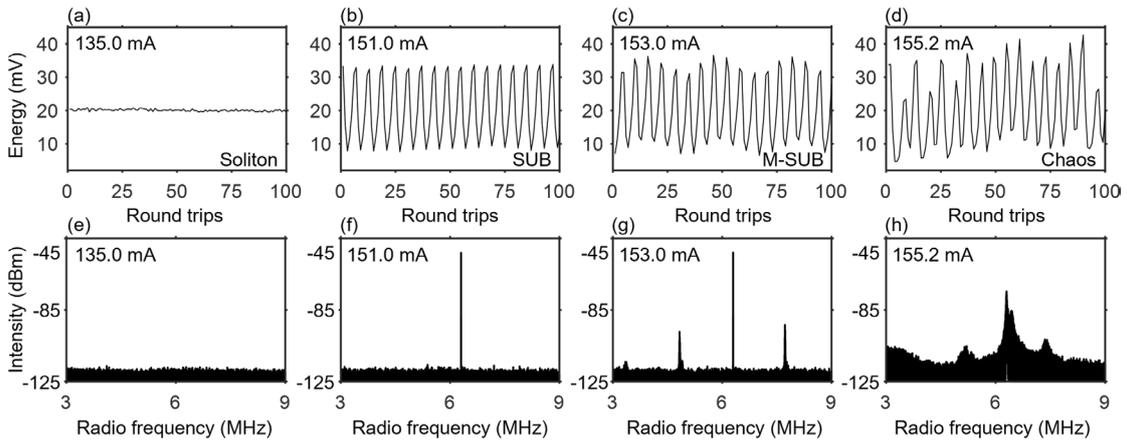

Fig. S2. Evolution of the pulse energy over cavity roundtrips for (a) stationary solitons (recorded at 135.0-mA pump current), and the (b) subharmonic (SUB) breathing soliton (151.0 mA), (c) modulated subharmonic (M-SUB) breathing soliton (153.0 mA), and (d) chaotic (155.2 mA) states illustrated in Fig. 2 of the main body of the manuscript. (e-h) Corresponding rf signals in the vicinity of $f_r/6$. Solitons have a nearly constant energy, while subharmonic breathing solitons feature periodic energy variations across a period of 6 cavity roundtrips. This periodicity manifests itself as a single spectral line (the breathing frequency, $f_b$) located at $f_r/6$ in the rf spectrum, while solitons do not exhibit any rf components in the considered frequency range. An additional modulation on the energy is apparent for the modulated subharmonic breathing soliton state, reflecting the emergence of two equally spaced sidebands on both sides of $f_b = f_r/6$ in the rf spectrum. The chaotic state features an irregular modulation on the energy and broad rf spectral lines.

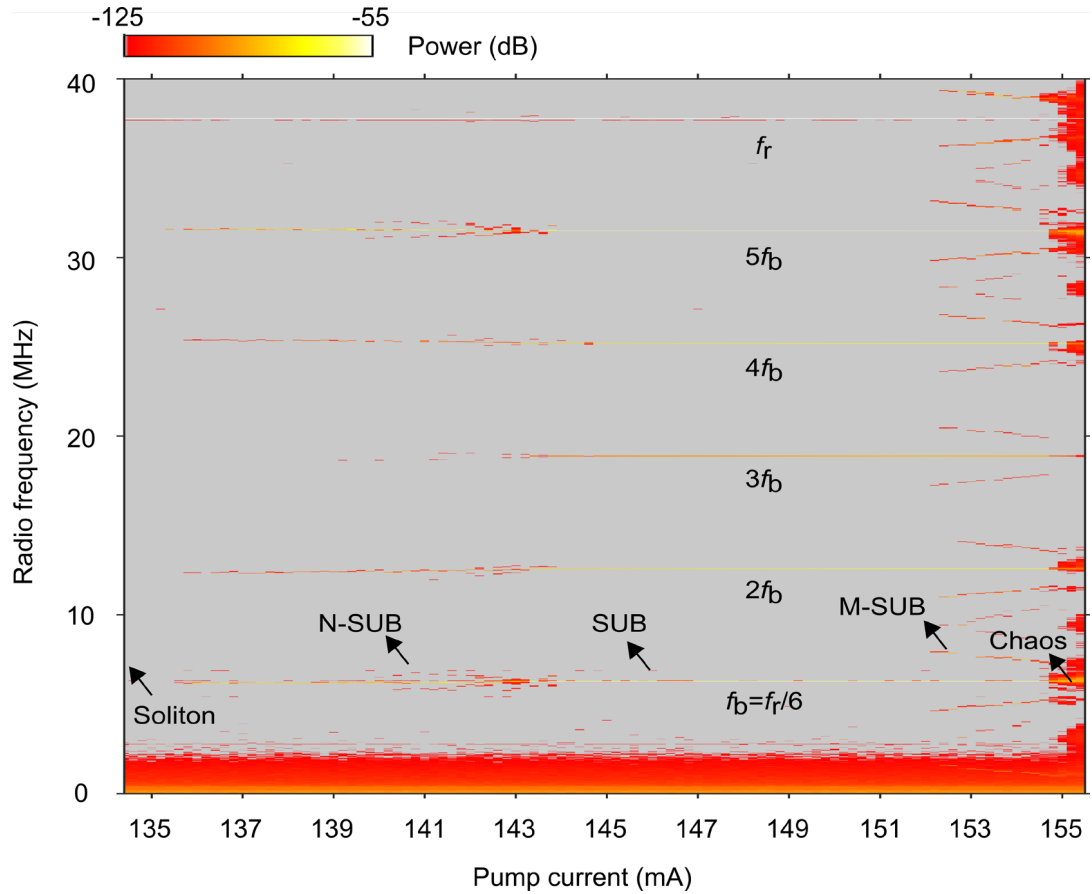

Fig. S3. Extended version of Fig. 2a in the main body of the manuscript, where the rf ranges from 0 to $f_r$.

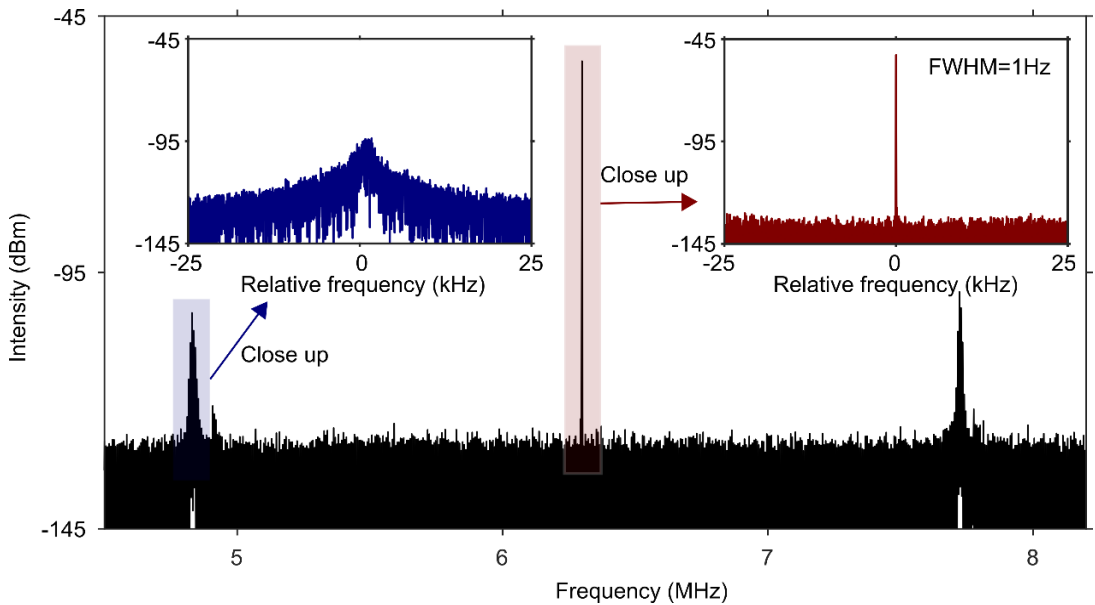

Fig. S4. Rf spectrum measurement for the modulated subharmonic breathing regime at a pump current of 153 mA, highlighting the instability of this state caused by the non-subharmonic sidebands. The left inset shows a broad and noisy structure of the sidebands, while the subharmonic breathing frequency shown in the right inset has a narrow linewidth (~1 Hz).

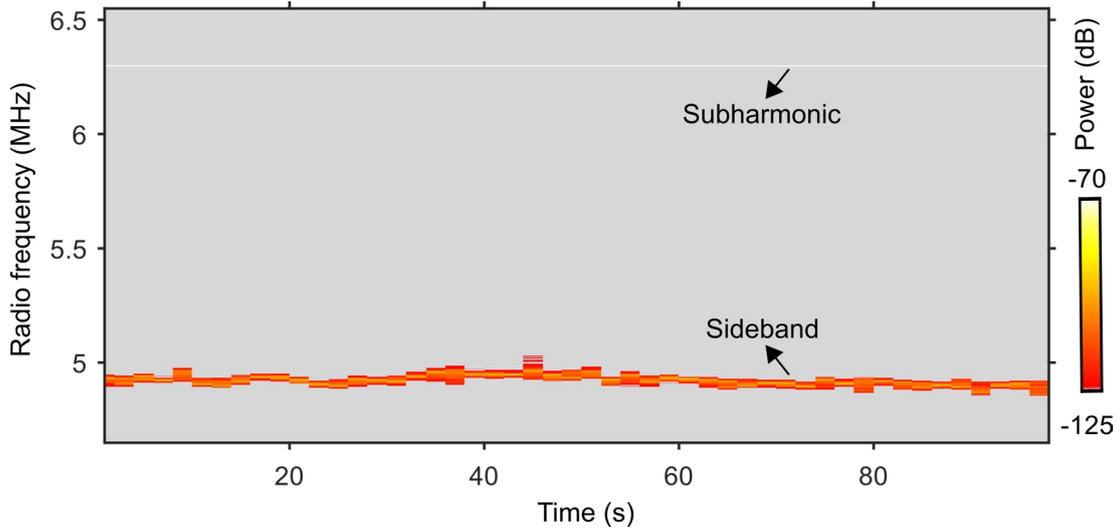

Fig. S5. Time evolution of the subharmonic frequency and one sideband for the modulated subharmonic breathing regime illustrated in Fig. S4, showing that the subharmonic frequency remains constant over time whilst the sideband is not stable. The resolution of the rf analyzer is 10 Hz.

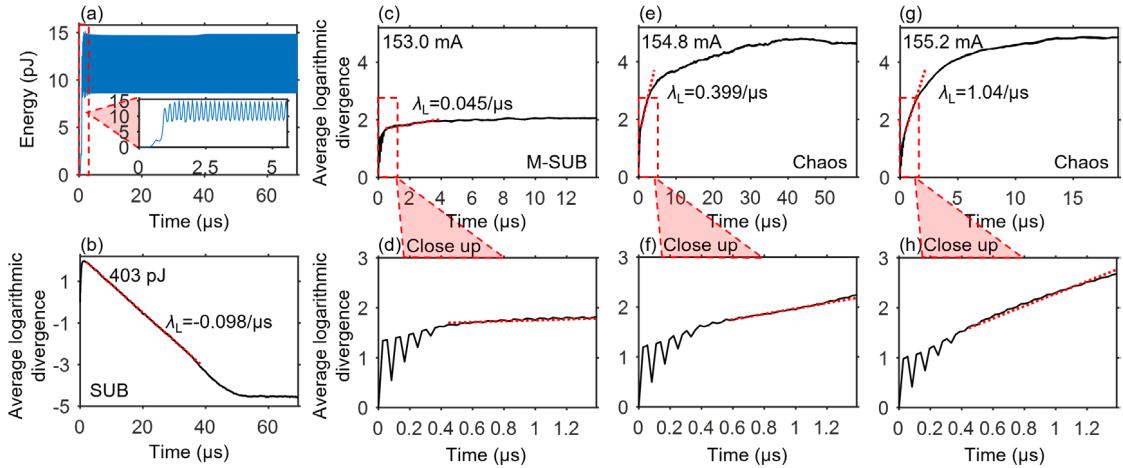

Fig. S6. (a) Evolution of the pulse energy over time showing the generation of subharmonic breathing solitons from the initial conditions. (b) Maximum Lyapunov exponent analysis for the subharmonic breathing state in (a), yielding the exponent fitting result $\lambda_L = -0.098/\mu s$ (red dotted line). Note that (a-b) show numerical simulation results as the subharmonic breathing regime is not self-starting in the experiment, namely, the evolution from initial conditions to the subharmonic state cannot be measured experimentally, which makes it difficult to calculate the associated Lyapunov exponent. (c-g) Maximum Lyapunov exponent analysis for the modulated subharmonic (153.0 mA) and chaotic (154.8 mA and 155.2 mA) states illustrated in panels II, III and IV, respectively, of Fig. 2 in the main body of the manuscript. (d-h) Close-up views of the exponential divergences in (d-h). Due to the inevitable noise in the measurements, the modulated subharmonic state features a positive $\lambda_L$, yet much smaller than those of the chaotic states.

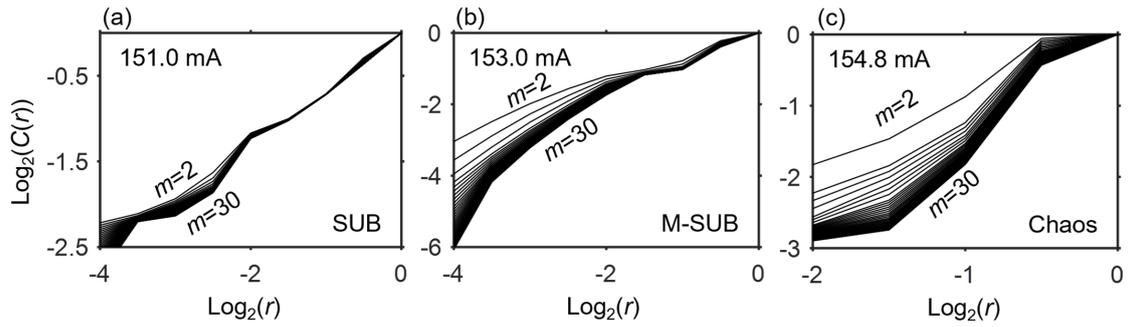

Fig. S7 (a-c) Plots of the correlation integral ($C(r)$) versus the sphere radius ($r$) for varying embedding dimension ($m$, from 2 to 30), for the subharmonic (151.0 mA), modulated subharmonic (153.0 mA) and chaotic resonance (154.8 mA) states illustrated in panels I, II and III, respectively, of Fig. 2 in the main body of the manuscript.

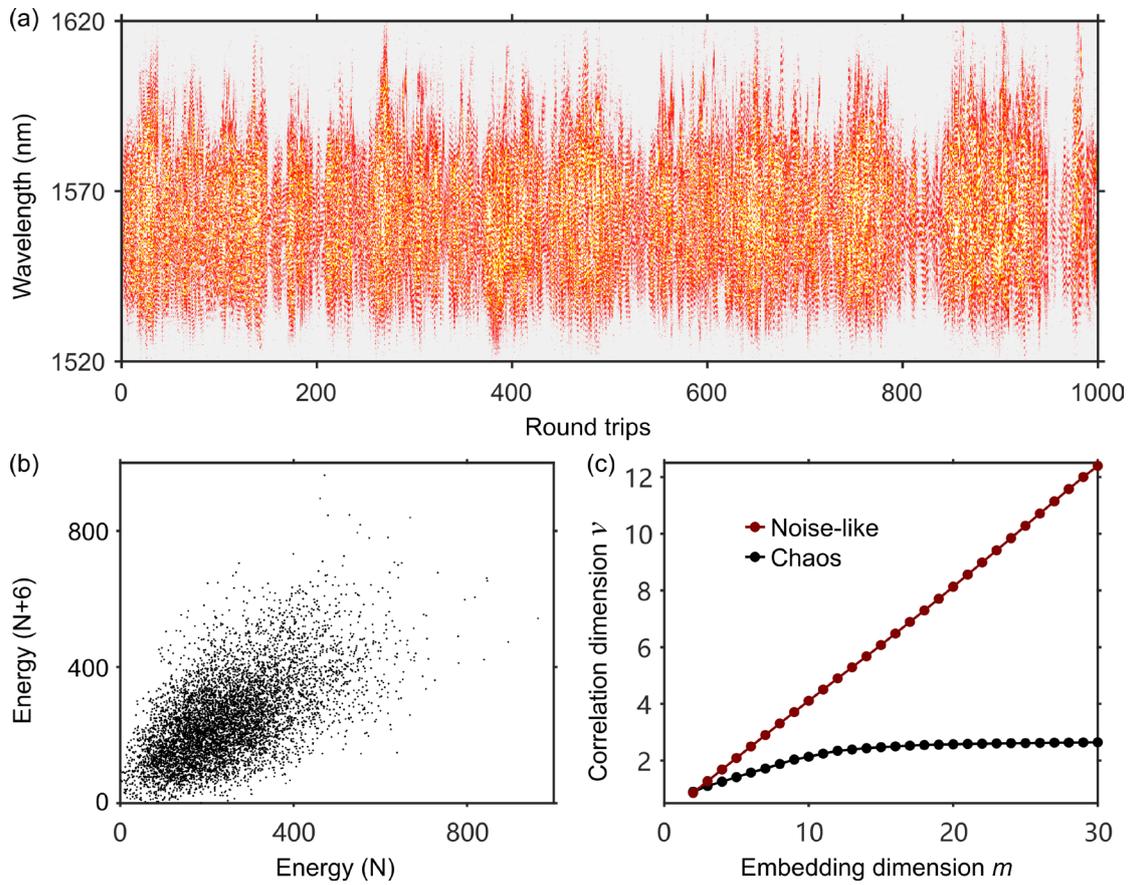

Fig. S8. Experimentally observed noise-like pulse operation of the laser. (a) Time-stretch recording of single-shot optical spectra over consecutive cavity roundtrips. (b) Poincaré section, showing the phase portrait of the pulse energy at roundtrip $N$ versus the energy at roundtrip ($N+6$), exhibiting a random structure. (c) Corresponding correlation dimension, showing linear increase with the embedding dimension (red), compared with the saturation behavior of the correlation dimension for the chaotic system (black).

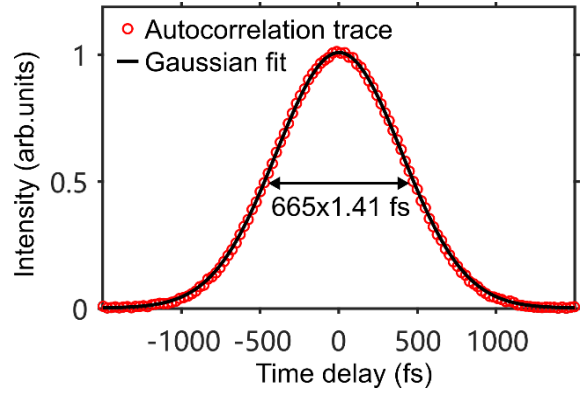

Fig. S9. Measured auto-correlation trace for the stable soliton state in Fig. 1b, c of the main body of the manuscript, showing a pulse duration of 665 fs (as estimated from a Gaussian fit), which is far beyond the resolution of the photodetector.

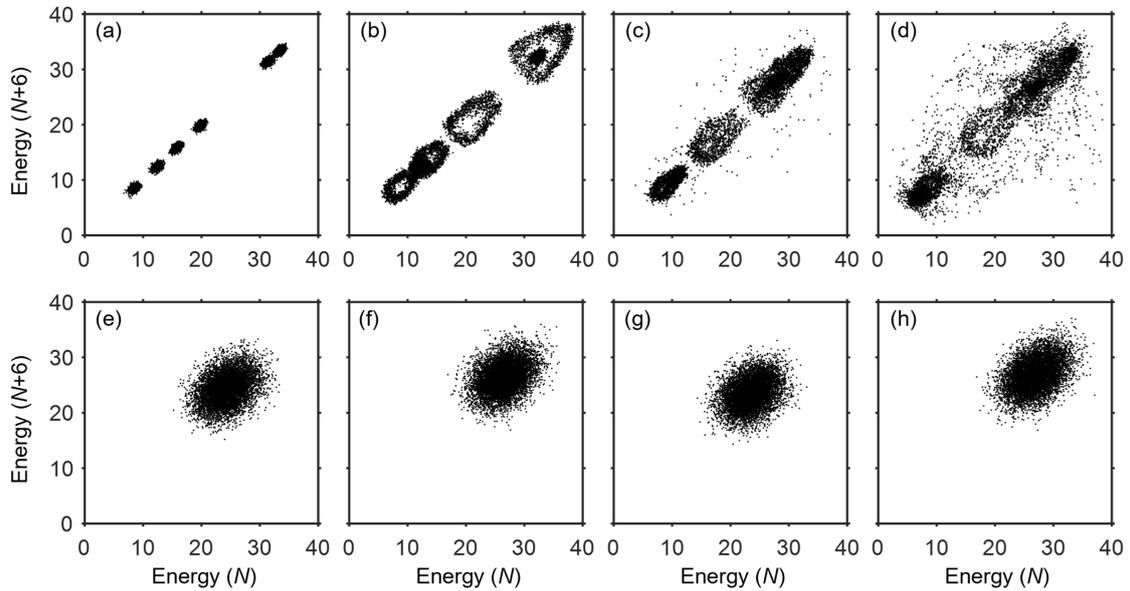

Fig. S10. (a-d) Poincaré maps shown in Fig. 2c of the main body of the manuscript, obtained by time stretch. (e-h) Respective Poincaré sections constructed by direct temporal intensity measurements without time stretching. The detailed structures of the attractors in (a-d) cannot be unveiled in (e-h) because of the limited temporal resolution of the photodetector, which evidences the importance of time stretch for exploring chaotic dynamics in mode-locked lasers.

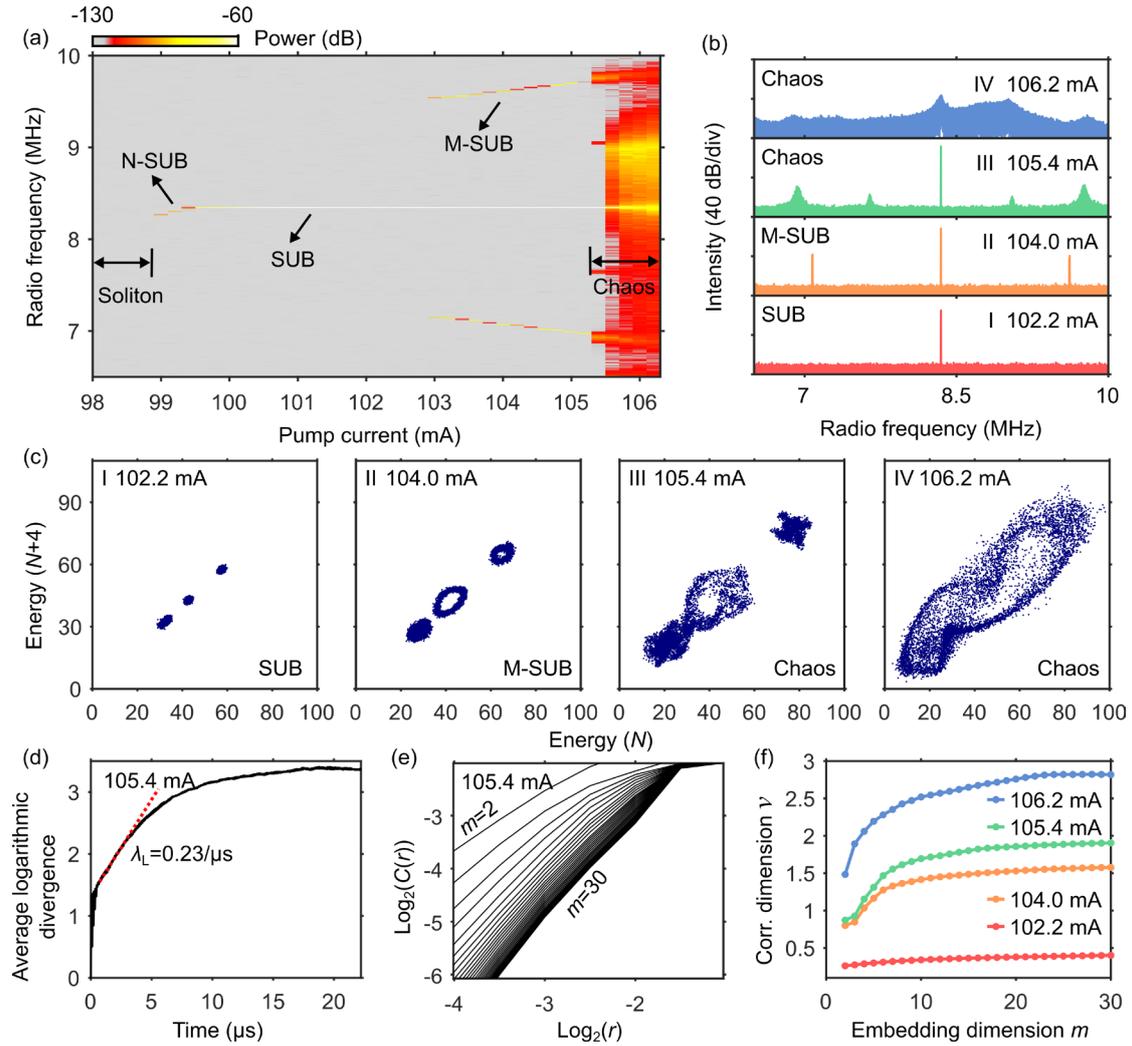

Fig. S11. Experimental observation of modulated subharmonic route to chaos for the ring-cavity laser shown in Fig. S1. (a) The laser output rf spectrum as a function of the pump current, showing successive phase transitions from stationary solitons to non-subharmonic (N-SUB), subharmonic (SUB), modulated subharmonic (M-SUB) breathing solitons and to chaos. (b) Representative rf spectra of subharmonic (102.2 mA), modulated subharmonic (104 mA), and chaotic (105.4 and 106.2 mA) states. (c) Corresponding Poincaré sections. (d) Maximum Lyapunov exponent analysis of the chaotic state in panel IV of (c), yielding the exponent fitting result $\lambda_L = 0.23/\mu s$. (e) Plots of the correlation integral ($C(r)$) versus sphere radius ($r$) for varying embedding dimension for the chaotic state in panel IV of (c). (f) Correlation dimension versus embedding dimension for the four dynamic states in (b).

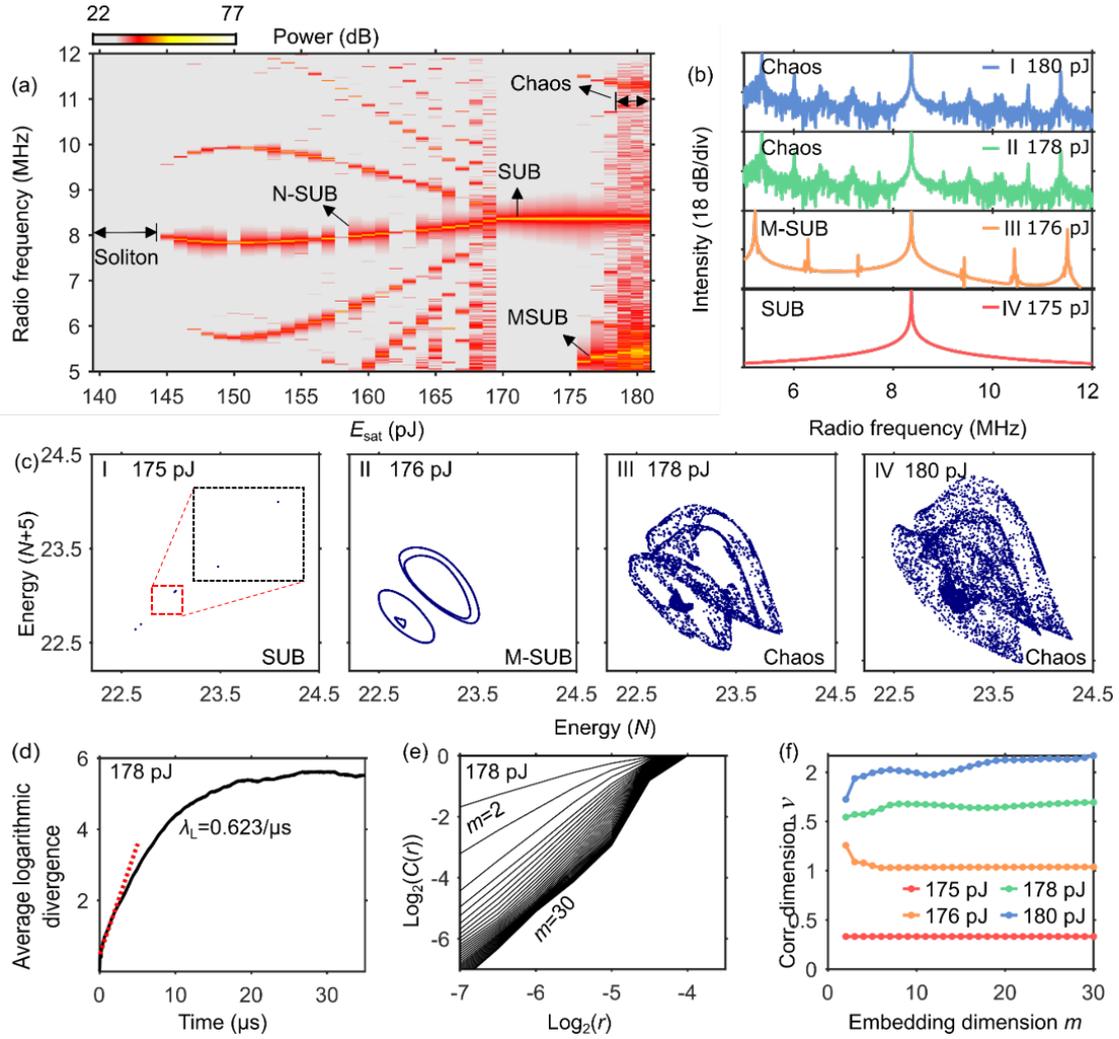

Fig. S12. Numerical simulation of modulated subharmonic route to chaos for the ring-cavity laser shown in Fig. S1. (a) Laser rf spectral intensity as a function of the gain saturation energy, showing transitions among different laser states. (b) Representative rf spectra of subharmonic (175 pJ), modulated subharmonic (176 pJ) and chaotic (178 and 180 pJ) states. (c) Corresponding Poincaré sections. (d) Maximum Lyapunov exponent analysis of the chaotic state in panel III of (c), yielding the exponent fitting result $\lambda_L = 0.623/\mu s$. (e) Plots of the correlation integral ($C(r)$) versus the sphere radius ($r$) for varying embedding dimension. (f) Correlation dimension versus embedding dimension for the four dynamic states in (b).

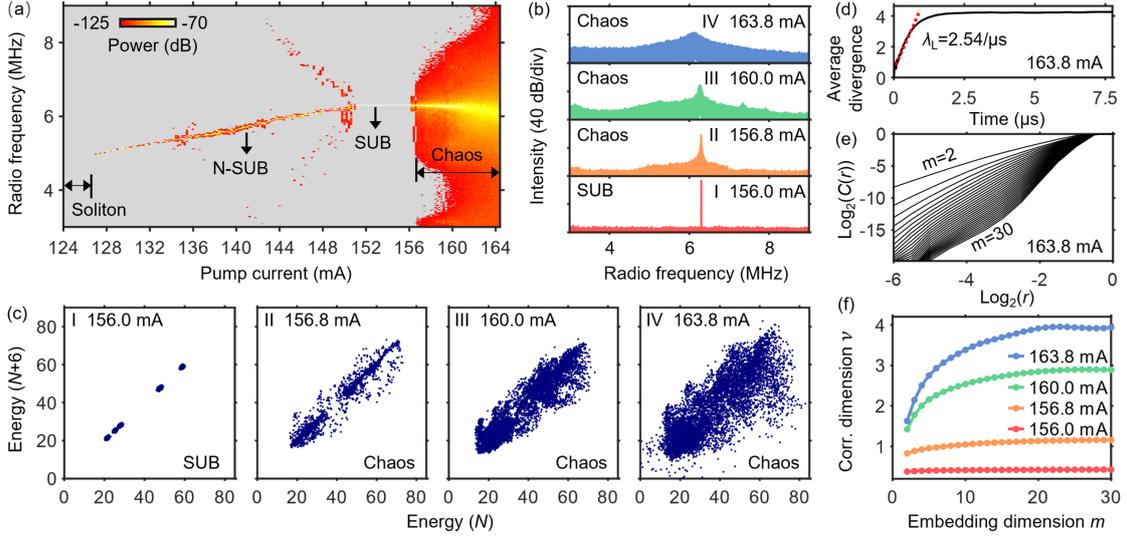

Fig. S13. Experimental observation of subharmonic route to chaos. (a) Spectrogram of the laser output rf spectrum as a function of the pump current, showing successive phase transitions from stationary solitons to non-subharmonic (N-SUB), subharmonic (SUB) breathing solitons, and to chaos. (b) Representative rf spectra of the subharmonic state (156.0 mA) and of chaotic states under different pump currents. (c) Corresponding Poincaré sections. (d) Maximum Lyapunov exponent analysis of the chaotic state in panel IV of (c), yielding the exponent fitting result $\lambda_L = 2.54/\mu s$. (e) Plots of the correlation integral ($C(r)$) versus sphere radius ($r$) for varying embedding dimension for the chaotic state in panel IV of (c). (f) Correlation dimension versus embedding dimension for the four dynamic states in (b).

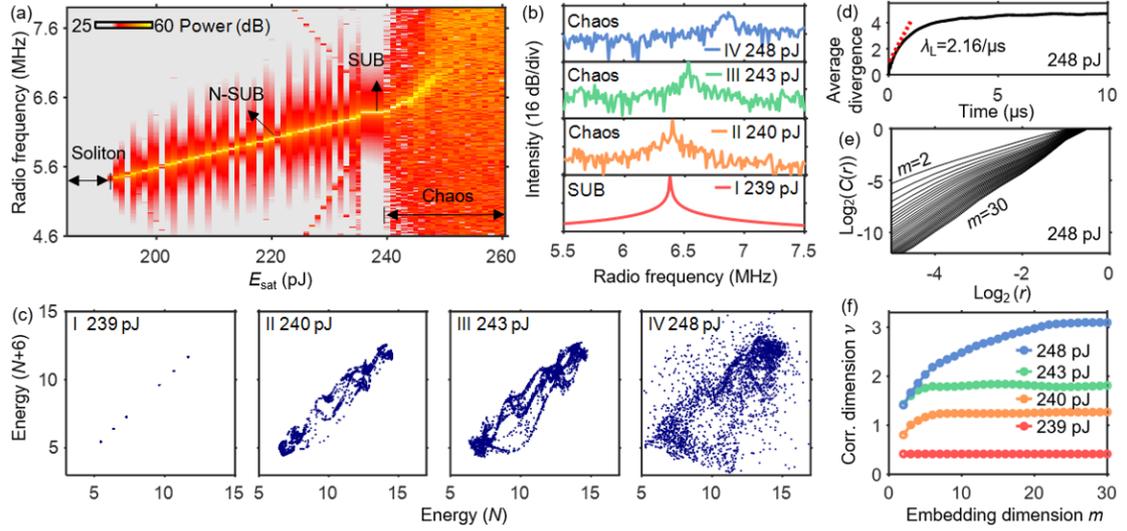

Fig. S14. Numerical simulation of subharmonic route to chaos. (a) Spectrogram of the laser rf spectral intensity as a function of the gain saturation energy, showing transitions among different laser states. (b) Representative rf spectra of the subharmonic state (239 pJ) and of chaotic states under different gain saturation energies. (d) Maximum Lyapunov exponent analysis of the chaotic state in panel IV of (c), yielding the exponent fitting result $\lambda_L = 2.16/\mu s$. (e) Plots of the correlation integral ($C(r)$) versus sphere radius ($r$) for varying embedding dimension for the chaotic state in panel IV of (c). (f) Correlation dimension versus embedding dimension for the four dynamic states in (b).

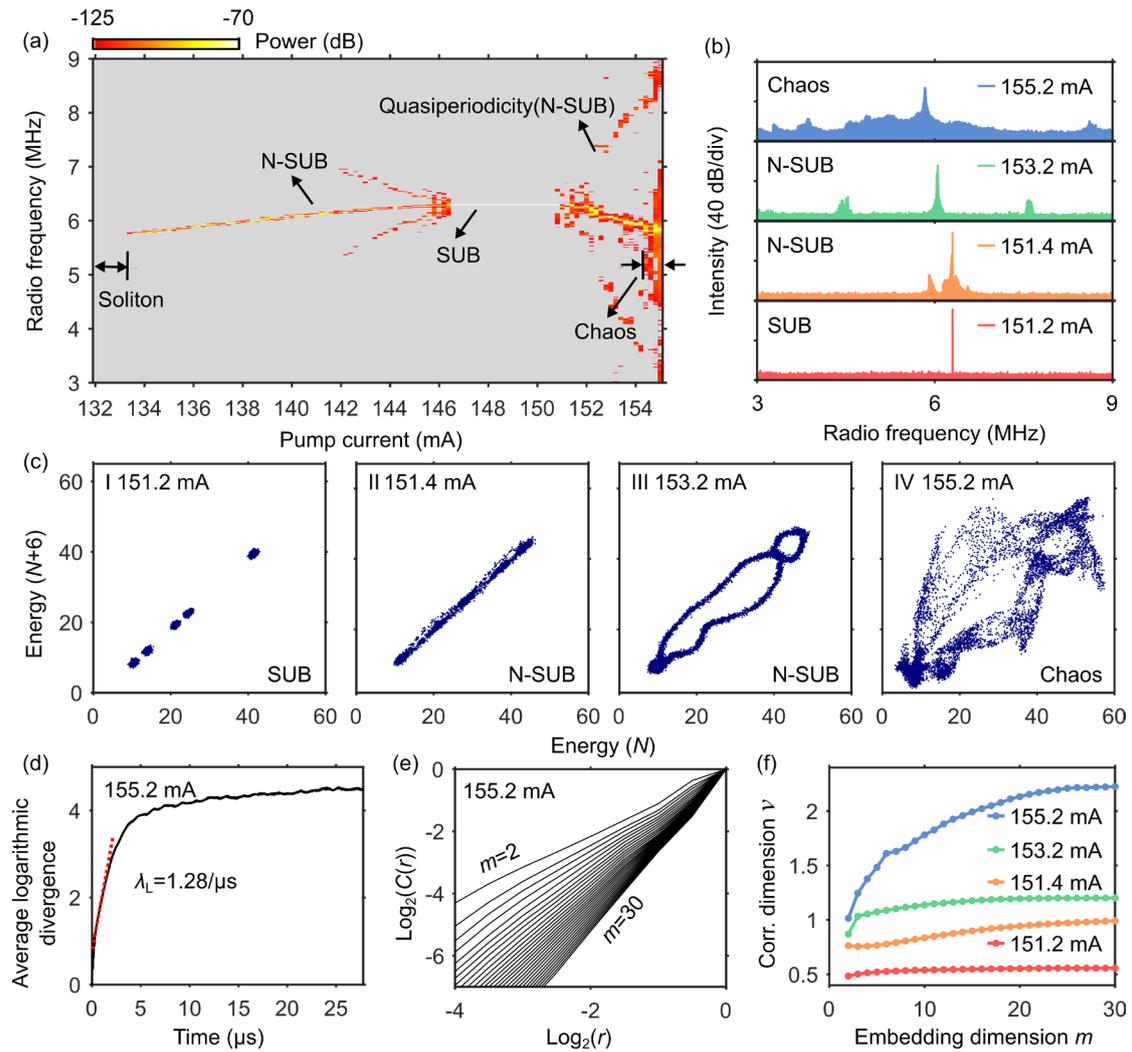

Fig. S15. Experimental observation of non-subharmonic (quasiperiodicity) route to chaos. (a) Spectrogram of the laser output rf spectrum as a function of the pump current, showing successive phase transitions from stationary solitons to non-subharmonic (N-SUB), subharmonic (SUB), non-subharmonic (N-SUB) breathing solitons and to chaos. (b) Representative rf spectra of subharmonic (151.2 mA), quasiperiodic (151.4 and 153.2 mA), and chaotic (155.2 mA) states. (c) Corresponding Poincaré sections. (d) (d) Maximum Lyapunov exponent analysis of the chaotic state, yielding the exponent fitting result $\lambda_L = 1.28/\mu s$. (e) Plots of the correlation integral ($C(r)$) versus sphere radius ($r$) for varying embedding dimension for the chaotic state. (f) Correlation dimension versus embedding dimension for the four dynamic states in (b).

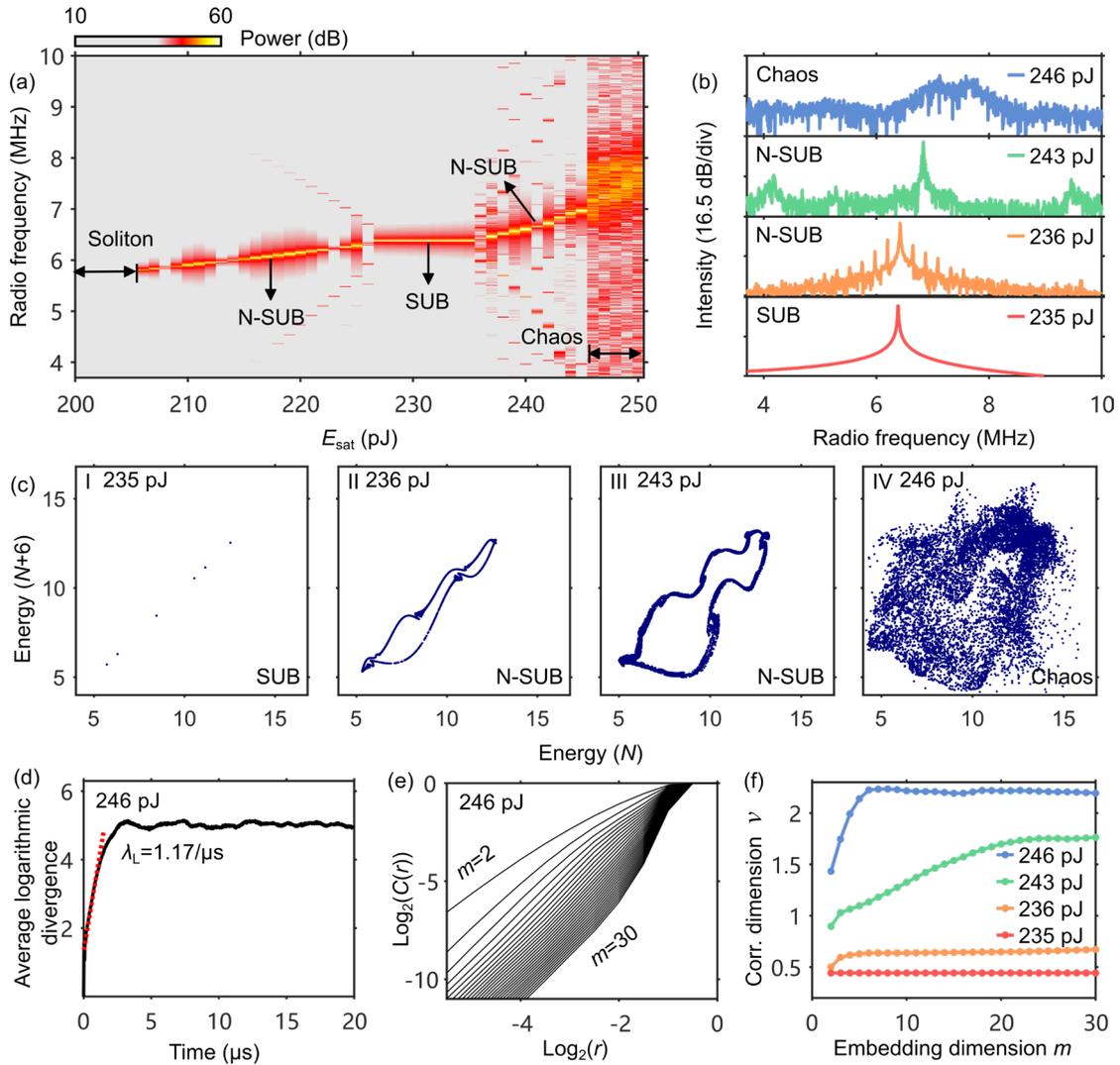

Fig. S16. Numerical simulation of non-subharmonic route to chaos. (a) Spectrogram of the laser rf spectral intensity as a function of the gain saturation energy, showing transitions among different laser states. (b) Representative rf spectra of subharmonic (235 pJ), quasiperiodic (236 and 243 pJ), and chaotic (246 pJ) states. (c) Corresponding Poincaré sections. (d) Maximum Lyapunov exponent analysis of the chaotic state, yielding the exponent fitting result $\lambda_L = 1.17/\mu s$. (e) Plots of the correlation integral ($C(r)$) versus sphere radius ($r$) for varying embedding dimension for the chaotic state. (f) Correlation dimension versus embedding dimension for the four dynamic states in (b).

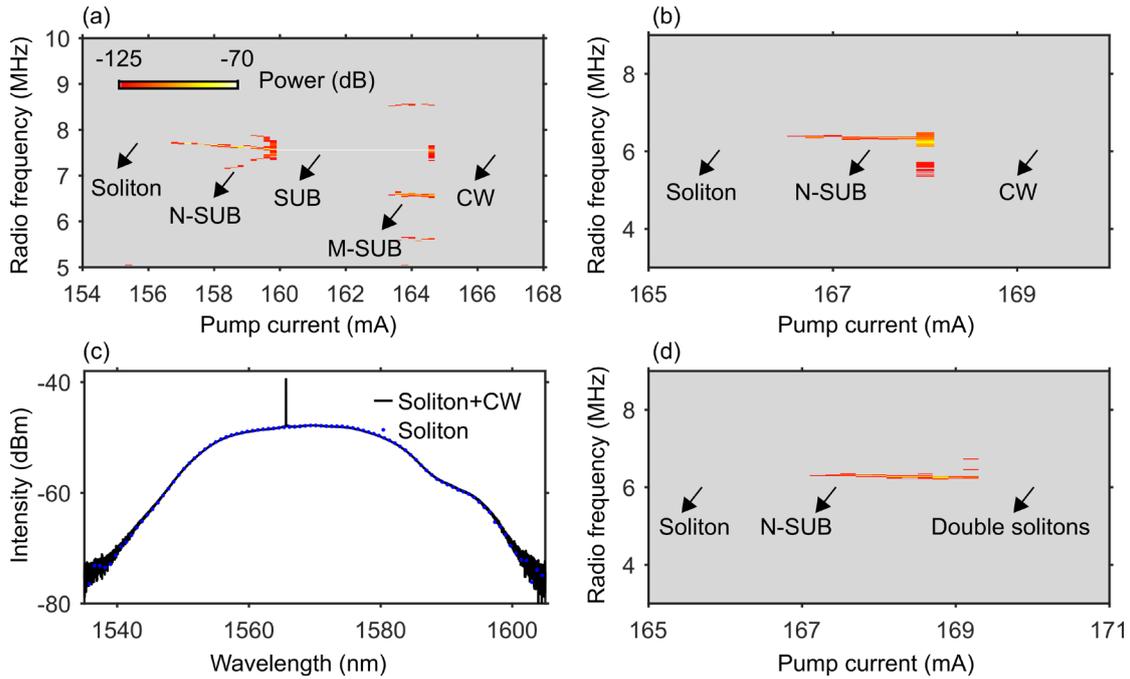

Fig. S17. Other scenarios in which solitons cease to exist in the laser when the pump current is increased. (a-b) Rf spectra of the laser output as a function of the pump current, showing that continuous waves (CWs) are generated when solitons cease to exist. In (a), successive phase transitions from stationary solitons to non-subharmonic (N-SUB), subharmonic (SUB) and modulated subharmonic (M-SUB) breathing solitons occur before CWs appear, while in (b) only the non-subharmonic breathing state is observed before the CW emergence. Note that no rf signature of the soliton mode of operation can be observed in the frequency range considered. (c) Optical spectra of the laser output showing that the soliton state (blue curve) is destabilized by the development of a spike (CW; black curve) on the soliton spectrum. (d) Rf spectrum of the laser output as a function of the pump current, showing that a soliton pair is generated from the soliton state after a transition through the non-subharmonic breathing state.

Table 1 Parameters used in the numerical model (figure-of-eight laser).

| Route to chaos | Output coupler | Phase delay | SMF (SMF28) | EDF (OFS 80) |
|---|---|---|---|---|
| Modulated subharmonic | Coupling coefficient $T_{OC} = 0.502$. This includes also other losses (fiber loss, splicing loss between fiber, and losses from the polarization controller). | $\theta = 0.538\pi$. This delay arises mainly from the polarization controllers in the bidirectional loop. | GVD coefficient $\beta_2 = -0.0173 \text{ps}^2/\text{m}$; nonlinearity coefficient $\gamma = 0.0011 \text{ (W}\cdot\text{m})^{-1}$; length $L = 4.14$ m. | $\beta_2 = 0.0619 \text{ ps}^2/\text{m}$; $\gamma = 0.01 \text{ (W}\cdot\text{m})^{-1}$; $L = 1.22$ m; gain bandwidth = 50 nm; small-signal gain $g_0 = 4.0/\text{m}$; gain saturation energy $E_{sat} = 290 - 420$ pJ. |
| Subharmonic | $T_{OC} = 0.485$ | $\theta = 0.57\pi$ | | $E_{sat} = 170 - 245$ pJ |
| Non-subharmonic | $T_{OC} = 0.42$ | $\theta = 0.595\pi$ | | $E_{sat} = 200 - 255$ pJ |

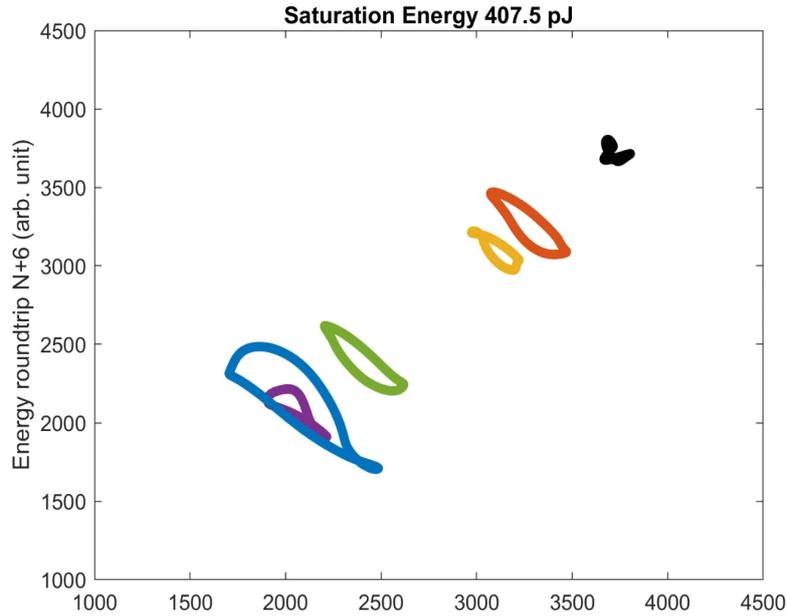

Video 1. This video shows how the phase diagrams evolve by varying the gain saturation energy in the modulated subharmonic route to chaos (referring to Fig.3c in the main text). A very small step size of gain saturation energy is employed to see the transitions between different phases (Double click to see the video).